\documentclass{amsart}
\usepackage[dvips]{graphicx}
\usepackage{amssymb,amsmath}

\begin{document}


\title{Discrete Scale Invariance in Scale Free Graphs}

\author{Mat\'\i as Gra\~na}
\address{MG: Departamento de Matematicas, FCEyN, Universidad de Buenos Aires,
Ciudad Universitaria, Pab. 1 (1428) Ciudad de Buenos Aires, Argentina.}
\email{matiasg@dm.uba.ar}
\thanks{MG is partially supported by  Fundaci\'on Antorchas and CONICET}
\author{Juan Pablo Pinasco}
\address{JPP: Instituto de Ciencias, Universidad de General Sarmiento,
J.M.Gutierrez 1150, (1613) Los Polvorines, Buenos Aires, Argentina.}
\email{jpinasco@ungs.edu.ar}
\thanks{JPP is partially supported by Fundaci\'on Antorchas and ANPCyT}

\begin{abstract}
In this work we introduce an energy function in order to study finite scale free
graphs generated with different models. The energy distribution has a fractal pattern
and presents log periodic oscillations for high energies. This oscillations are
related to a discrete scale invariance of certain graphs, that is, there are preferred
scaling ratios suggesting a hierarchical distribution of node degrees. On the other
hand, small energies correspond to graphs with evenly distributed degrees.
\end{abstract}

\maketitle

\bigskip
\section{Introduction}

In the last few years a huge amount of research on graphs has been achieved in physics
and mathematics, mainly due to the increasing importance of complex networks like the
Internet and the World Wide Web, and the role of social networks in the propagation of
diseases, infections, rumors and news, in both real and virtual environments such as
populations or web-based communities.

Despite the different origins, methods and perspectives in those works, certain
concepts appeared and pervaded the network literature, like small worlds, clustering,
centrality, scale free degree distribution, and preferential attachment, among others,
\cite{ASBS}, \cite{BA}, \cite{PN}. They suggest that simple underlying mechanisms could
describe and explain the network growth and their emerging properties. To this end,
several networks based on real data were analyzed (see \cite{DM}, \cite{Nepareto} and
the references therein), and many theoretical models of network growth were proposed
(see Table 3 in \cite{BA}).

Also, there are still many questions and problems which remain unanswered. Perhaps the
main one is the origin of the power law degree distribution of nodes in real networks.
This distribution seems to be observed in several real networks (although some
criticism appeared in the last year,  \cite{ACKM}, \cite{GMY}, \cite{Mi}). Roughly
speaking, the power law distribution says that the number of nodes in a graph with $k$
links for $k>k_0$ has a decay proportional to $k^{-\alpha}$, for certain $\alpha
>1$. Several theoretical models were proposed to explain this phenomenon. Among them,
let us mention \emph{preferential attachment} of Barabasi and Albert \cite{AB},
\emph{edge redirection} of Krapivsky and Redner \cite{KR}, and \emph{attachment to
edges} of Dorogovtsev, Mendes and Samukhin \cite{DMS}.

Although much attention was paid in the theoretical literature in the thermodynamical
limit (when time and number of nodes tend to $\infty$) in these models, not too much
is known for graphs with a bound in their size (\cite{DMS} being an interesting
exception). Indeed, many of the computer-based simulations tend to concentrate on
large graphs, while small ones usually get ignored. Yet, real networks are finite, and
in many cases they are very small (like networks of metabolic reactions, proteins
interactions, digital electronic circuits, sexual interactions or food webs). Our
intention in this paper is to concentrate on these small graphs. Specifically, we
concentrated on graphs with a number of nodes ranging from 50 to 100,000. On the other
hand, we were able to generate large amounts of graphs for each model, which reduced
the noise and statistical fluctuations, and helped see some otherwise hidden
properties.

Another question, which motivates our research, was the discrimination among the
different graphs generated from these models and the analysis of their properties. We
may consider this question as a basic one concerning the internal topology of the
network, and the quality of their connections. Let us note that different networks
with the same node distribution respond in distinct ways to edges or nodes removal
depending on the connectivity between hubs, that is, if the nodes with high number of
links are connected or not between them. This question was considered by Newman in
\cite{Ne}, who coined the term assortativity in order to describe this phenomenon.

In order to study this question, we introduce the linking energy of a graph $G$,
defined as
$$E_2(G) = \sum_{x\sim y} |d(x)-d(y)|^2,$$
where $d(x)$ is the degree of the node $x$, i.e., its number of links, and the sum
runs over the pairs of connected nodes.

Our original intention was to compute the expected linking energy for certain
scale-free network generating models, which enabled us to classify the graphs
generated for each models as either assortative or not.  One of the first
observations, was that the underlying mechanism responsible of networks generation in
each model could generate both assortative and non-assortative networks. Another
interesting fact is the correlation of the assortativity and the linking energy $E_2$,
and the fractal structure of this correlation. These results will be published in a
separated work.

However, we also obtained an unexpected result when we computed $E_2$ for the models
mentioned above, since the distribution of energies presents log-periodic oscillations
in the tail. This oscillatory behavior is more apparent and becomes amplified within
models which tend to produce a low number of highly connected hubs. The oscillatory
behavior is not present for random or regular graphs, even when they are small.

Surprisingly, this shed some light on the first problem, the node distribution. The
presence of log-periodic oscillations is associated to fractal complex dimensions,
which are due to discrete scale invariance instead of a scale free similarity
(\cite{SJAMS}, \cite{S}). Indeed, this suggests a stronger order in the networks other
than that predicted by the scale free models, since fractal complex dimensions are
related to (mathematical) self similar fractals such as the triadic Cantor set.
Hence, we may classify the graphs generated with any of those models as continuous or
lacunary scale free. A similar dichotomy is present for fractals generated with
iterated functions systems \cite{L-V}.

We conjecture that the nodes distribution itself must also exhibit the oscillations.
Further work is required to settle this question. However, several real networks that
appeared in the literature seem to show such oscillations, like the World Trade Web
\cite{SeBo}, or web-based communities (see Fig. 1, 5 and 8 in \cite{BL}). Also, two
theoretical models of networks without growth were recently proposed in \cite{XZW} and
\cite{BDS}, where oscillations are apparent.  The existence of oscillations on the
degree distribution was related to the existence of different hierarchies among the
nodes of the graph, see \cite{BR}, \cite{SH} where theoretical models were proposed
to study this kind of networks. Hence, our results show that classical models based on
preferential attachment or edge redirection can be used to generate non homogeneous
networks.

\bigskip

\section{Main Results}

\medskip
\subsection{The linking energy}

Let $G$ be any graph, and let us denote $x \sim y$ whenever there exists a link between
the nodes $x$ and $y$. Let $d(x)$ be the degree of $x$, that is, the number of links
attached to $x$.

We define the {\it linking energy} of $G$ as
$$E_2(G) = \sum_{x\sim y} |d(x)-d(y)|^2.$$
This is a discrete version of the energy which arises for the tension of a system of
strings attached at the nodes, with heights proportional to the number of connections.
For regular networks, it coincides with the variational form of the Dirichlet energy of
a discrete Laplacian.

Let us note that the energy spectra for graphs with $N$ nodes is finite. For a tree on
$N$ nodes, the maximum value of the energy is attained for the central tree where all
the nodes are connected to one of them, with an energy of
$$(N-1)(N-2)^2 \sim N^{3}.$$

Let us note as well that the energy is an even number:
$$
E_2(G)\equiv\sum_{x\sim y}d_x-d_y
\equiv\sum_{x\sim y}d_x+d_y
=\sum_xd_x^2\equiv\sum_xd_x=\sum_{x\sim y}2\equiv 0
\pmod 2
$$

For regular lattices, such as the triangular, square or hexagonal lattice, this energy
is proportional to the number of boundary nodes, since the degree is the same for all
the internal nodes. Also, a random (Erd\H os-Renyi) graph has lower energy than a
scale free one due to the small number of hubs in it.

For networks with power law degree distribution, this degree sequence is not enough to
characterize several topological and dynamical properties. Graphs with the same degree
sequence behave differently, depending on whether or not the hubs are connected among
them, a problem considered by Newman who introduced the {\it assortativity} to measure
the connectivity between hubs. The linking energy could be used in order to study this
problem.

\medskip
\subsection{$E_2$ in a random (Erd\H os-Renyi) graph}
Let $G$ be a graph in the Erd\H os-Renyi model, with $d$ nodes. Here, two nodes
connect each other with probability $p$. Let us fix an order for the nodes, say
$x_1,\cdots,x_d$, and for $1\le i<j\le d$ let $X_{ij}=\#\{k\sim i,\;k\neq j\}$ and
let $Y_{ij}=\#\{k\sim j,\;k\neq i\}$. In other words, $X_{ij}$ is the degree of the
node $i$ disregarding a possible connection to node $j$, and similarly for
$Y_{ij}$.
Then, $X_{ij}$ and $Y_{ij}$ are binomial, with probability
\[
p(X_{ij}=k) = p(X_{jj}=k) = \binom{d-2}k p^k(1-p)^{d-2-k}.
\]
Thus, the mean and the variance are given by
$$<X_{ij}>=p(d-2),$$
$$\sigma^2_{X_{ij}}=<X_{ij}^2>-<X_{ij}>^2=(d-2)p(1-p).$$

Furthermore, the variables $X_{ij}$ and $Y_{ij}$ are independent. Hence,
\begin{align*}
    <(X_{ij}-Y_{ij})^2> &= 2<X_{ij}^2> - 2<X_{ij}>^2 \\
        &= 2(d-2)p(1-p).
\end{align*}
Now, the mean value of $E_2$ is given by
\begin{align*}
<E_2> & = \sum_{i<j}p<(X_{ij}-Y_{ij})^2> \\
   &  = \binom d2p\cdot 2(d-2)p(1-p) \\
    & = d(d-1)(d-2)p^2(1-p).
\end{align*}

For other models of graphs, the distribution of energies gives better information than
the mean value of $E_2$.

\medskip
\subsection{Models of graph generation}

We explain now the studied models. All of them consist of the following steps:
\begin{itemize}
    \item One begins with a fixed seed of nodes and links.
    \item At each step, a new node is added to the graph, and
    \item a fixed number of links emanate from the new node to the existing ones.
\end{itemize}

These models differ in the way the targets for the new nodes are chosen. We considered
the following models:

\begin{itemize}
    \item[BA] \textbf{Preferential attachment} (or \emph{linear} P.A.): one target
        node is selected at random, with a probability proportional to the degree of
        the nodes.
    \item[KR] \textbf{Edge redirection}: one target node is selected at random with
        uniform probability. However, with probability $1-r$, this target is changed by
        the node it points to.
	\item[DMS] \textbf{Attach to edges}: one link is selected at random with uniform
		probability (among the links). The new node connects to both ends of the chosen
		link.
\end{itemize}
The labels are put after the authors of these models, see \cite{AB}, \cite{KR},
\cite{DMS}.

\medskip
\subsection{Numerical Experiments}

We computed the energy $E_2$ for several graphs constructed according to these models.
We present in this subsection some of these computations. For the KR model, we used
$r=0.5$, which gives a power law with exponent $3$ for the degree distribution, the
same exponent of the other two models (see \cite{BA}).

\begin{figure}[ht]
    \begin{center}
        \includegraphics[scale=0.9]{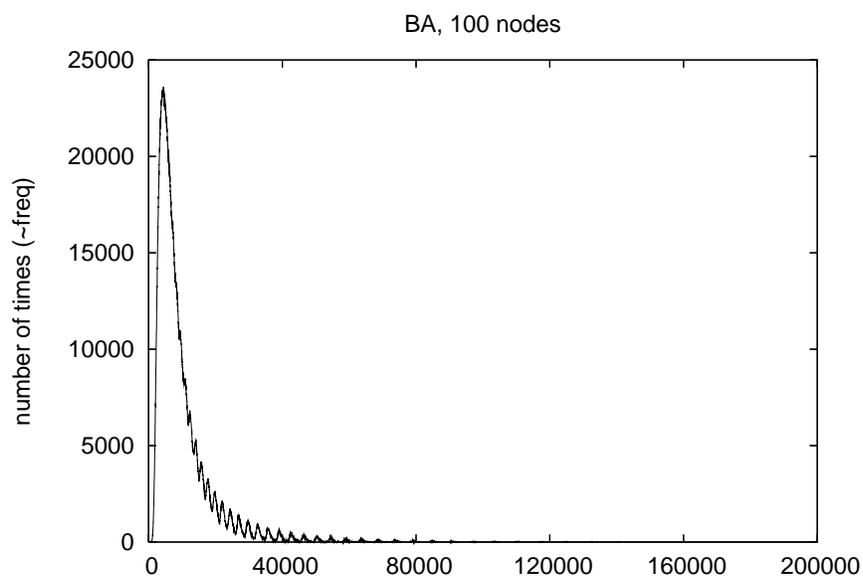}
    \end{center}
    \caption{Preferential attachment, $100$ nodes, histogram over $10^7$ graphs}
    \label{fig:pa1}
\end{figure}

\begin{figure}[ht]
    \begin{center}
        \includegraphics[scale=0.9]{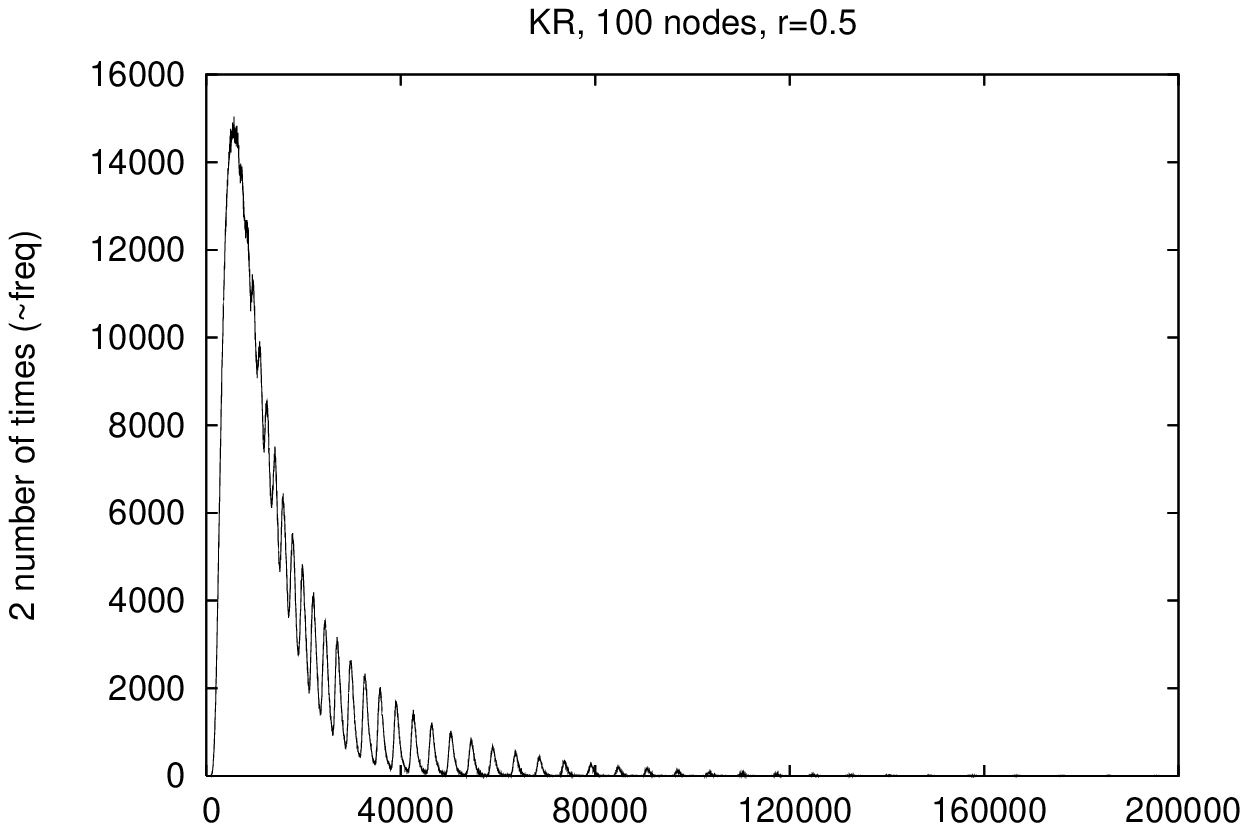}
    \end{center}
    \caption{Edge redirection, $100$ nodes, $r=0.5$, histogram over $10^7$ graphs}
    \label{fig:er1}
\end{figure}

\begin{figure}[ht]
    \begin{center}
        \includegraphics[scale=0.9]{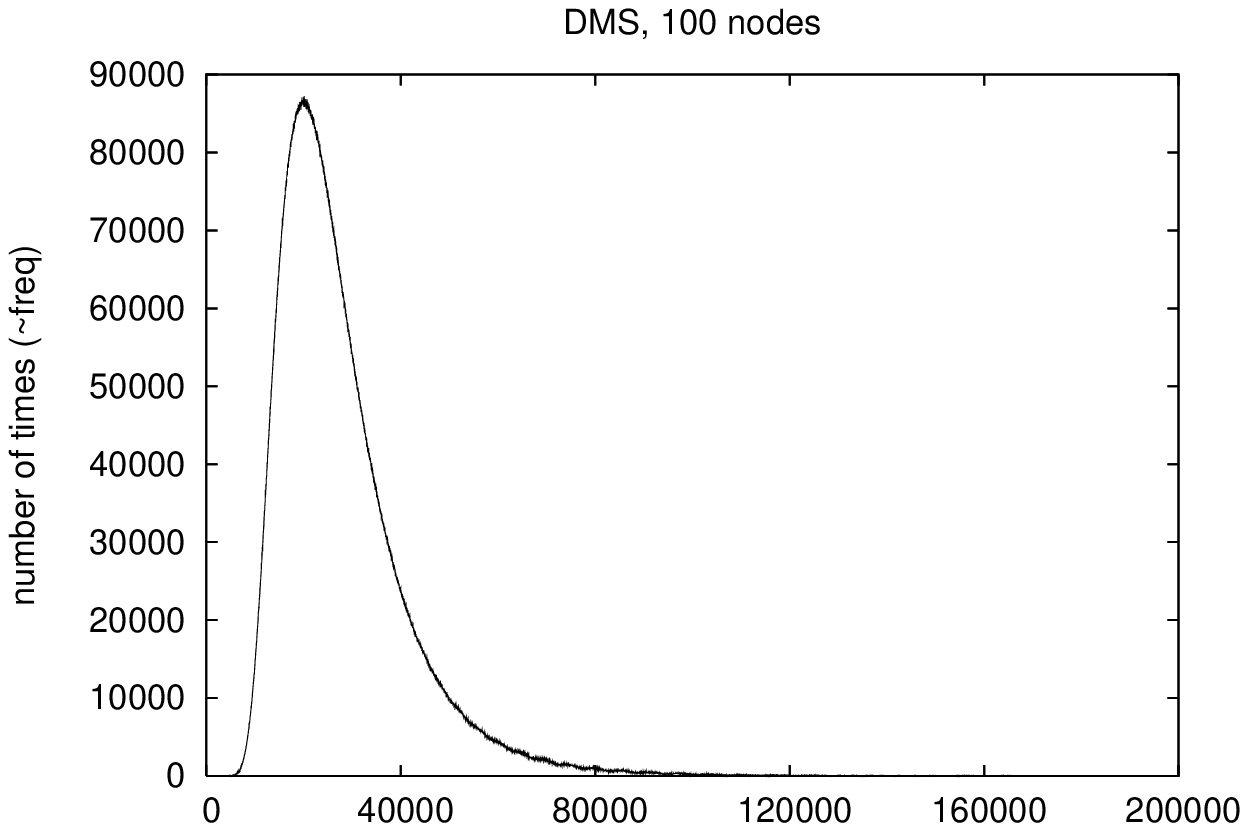}
    \end{center}
    \caption{Attach to edges, $100$ nodes, histogram over $10^8$ graphs}
    \label{fig:ate1}
\end{figure}

\begin{figure}[ht]
    \begin{center}
        \includegraphics[scale=0.9]{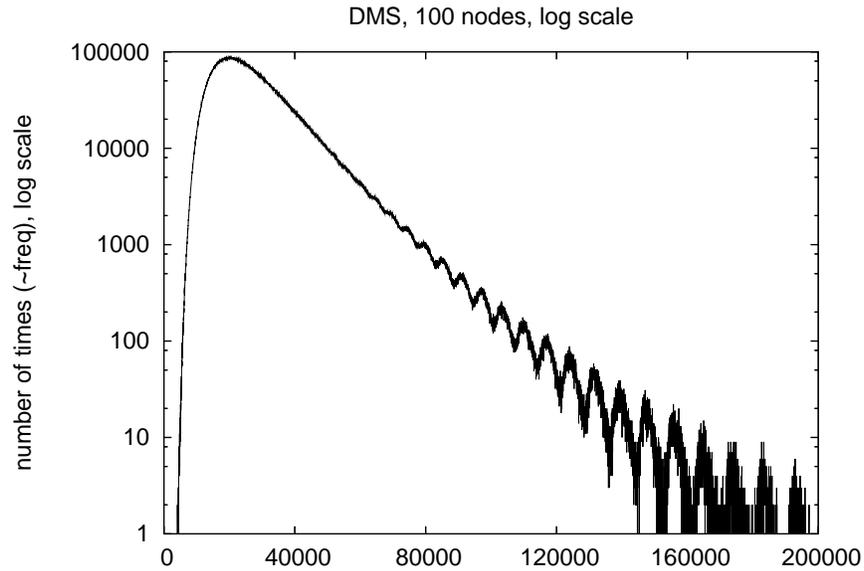}
    \end{center}
    \caption{Attach to edges, $100$ nodes, histogram over $10^8$ graphs, logarithmic
    scale in $E_2$}
    \label{fig:ate2}
\end{figure}

We present the results in Figures \ref{fig:pa1} to \ref{fig:ate2}:
in Figure \ref{fig:pa1} we show $E_2$ for $10^7$ graphs on $100$ nodes produced with
BA model. In Figure \ref{fig:er1} we show $E_2$ for $10^7$ graphs on $100$ nodes,
produced with KR model, with $r=0.5$. In Figure \ref{fig:ate1} we show $E_2$ for
$10^8$ graphs on $100$ nodes, produced with DMS model. Oscillations in this case become
more apparent using logarithmic scale, shown in Figure \ref{fig:ate2}.


Here, the energy spectra has a maximum close to $100^3$, although the observed
energy levels are far from this value in the examples. Moreover, there are no
forbidden levels at least for $E_2 \le 40,000$. This fact also shows that the
oscillations are not caused by the discreteness nor the sparsity at high values of the
energy.

\begin{figure}[ht]
    \begin{center}
        \includegraphics[scale=0.9]{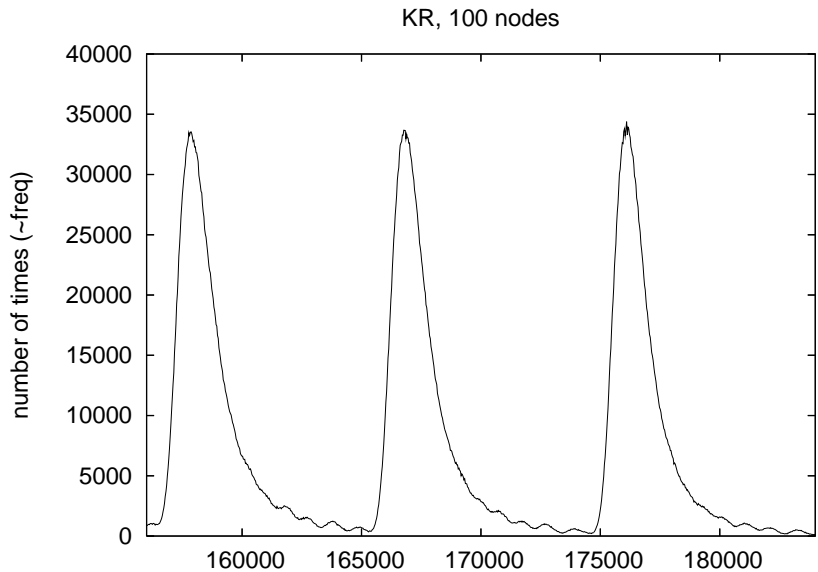}
    \end{center}
    \caption{Edge redirection, $100$ nodes, $r=0.25$, histogram over $10^8$ graphs}
    \label{fig:sub2}
\end{figure}

The oscillations persist even varying the width of the bins in the histogram. This
could be caused by the fractal pattern in the energy distribution, since each
oscillation exhibits a strong self-similarity with a sequence of
sub-os\-cil\-la\-tions, see
Figure \ref{fig:sub2}.

It is interesting to note that the three models shown above share the same node-degree
distribution. In all of the cases, degrees decay as a power-law of exponent $3$.
However, the energy distribution presents some differences. For instance, it is clear
that their modes differ: they are close to 4400 (BA), 5700 (KR), and 20000 (DMS).
Their oscillations also differ both in the place where they begin and in their
amplitudes.

\begin{figure}[ht]
    \begin{center}
        \includegraphics[scale=0.9]{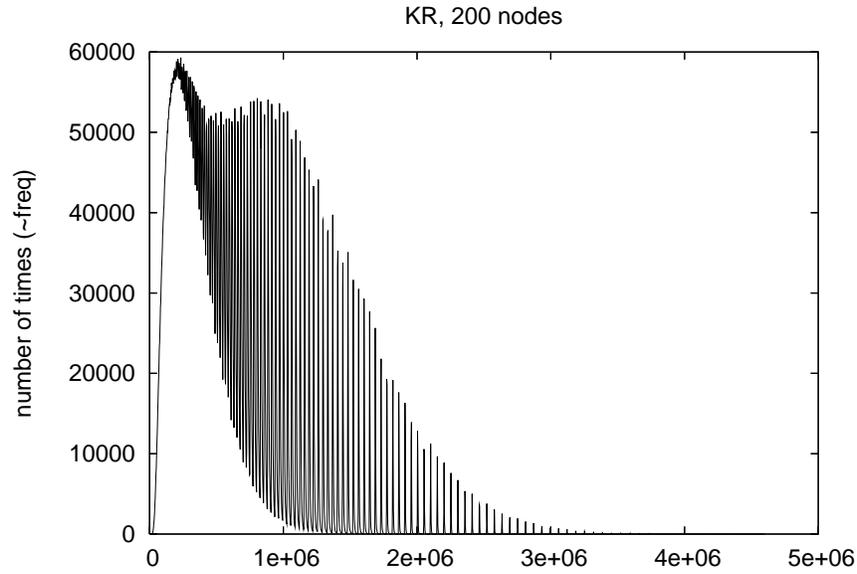}
    \end{center}
    \caption{Edge redirection, $200$ nodes, $r=0.25$, histogram over $2\times 10^7$ graphs}
    \label{fig:sub1}
\end{figure}

Also, for $r$ close to zero, the KR model gives high energy graphs, together with an
interesting behavior of the oscillations, which become highly amplified.
We conjecture that a phase transition occurs for some specific value $r=r_c$, a
phenomenon similar to that in \cite{LS}, which deserves further attention.
We show in Figure \ref{fig:sub1} the energy distribution for $r=0.25$.

\begin{figure}[ht]
    \begin{center}
        \includegraphics[scale=0.9]{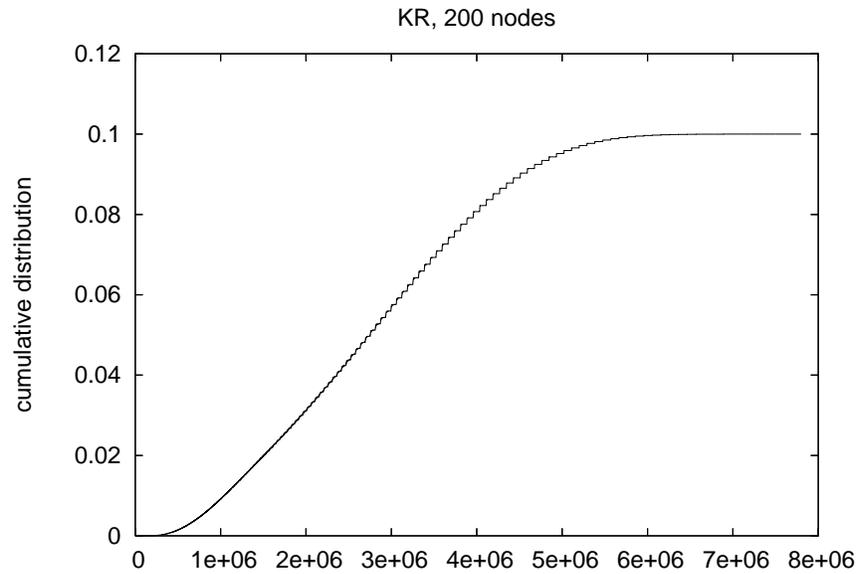}
    \end{center}
    \caption{Edge redirection, $200$ nodes, $r=0.25$, $10^7$ graphs}
    \label{fig:cantor1}
\end{figure}

\begin{figure}[ht]
    \begin{center}
        \includegraphics[scale=0.9]{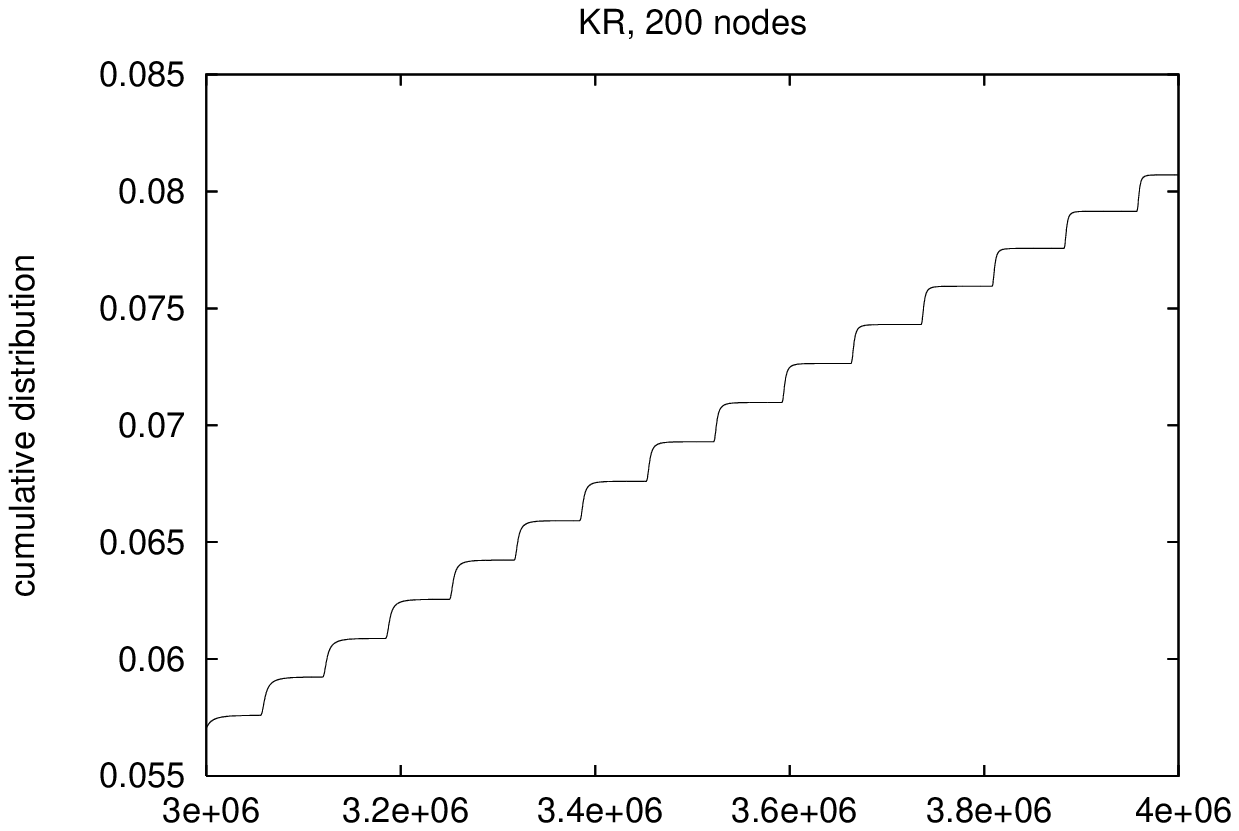}
    \end{center}
    \caption{Edge redirection, $200$ nodes, $r=0.25$, $10^7$ graphs}
    \label{fig:cantor2}
\end{figure}

\begin{figure}[ht]
    \begin{center}
        \includegraphics[scale=0.9]{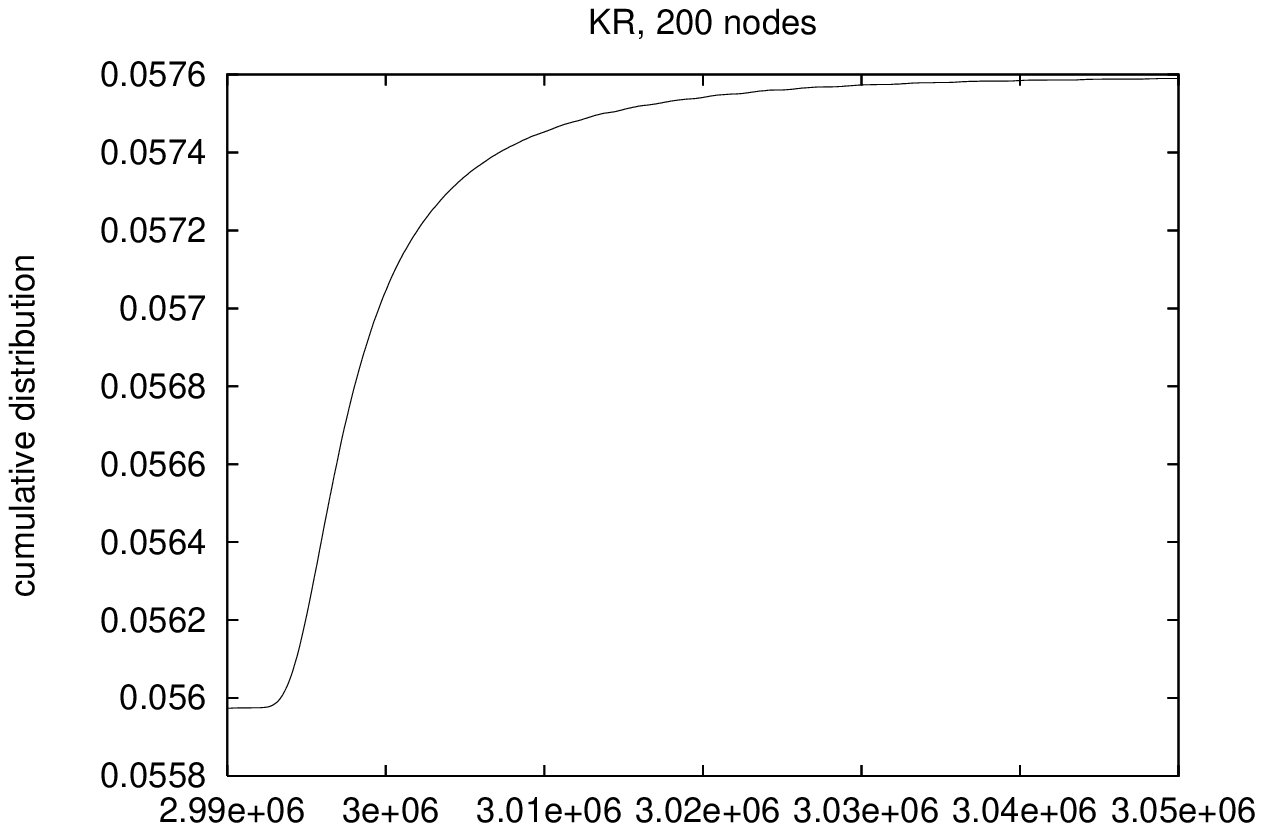}
    \end{center}
    \caption{Edge redirection, $200$ nodes, $r=0.25$, $10^7$ graphs}
    \label{fig:cantor3}
\end{figure}

The fractal pattern is clearly shown in the integrated density of energies,
$$N_2(E) = \int_0^e d\delta(E_2-E).$$
As was pointed out in \cite{Nepareto}, the analysis of the cumulative distribution is
better than binning in order to avoid statistical fluctuations. In Figure
\ref{fig:cantor1}, Figure \ref{fig:cantor2}, and Figure \ref{fig:cantor3}, we show
different amplifications of $N_2(E)$ for KR with $r=0.25$. Clearly, the structure
resembles a Cantor devil staircase.


\medskip
\subsection{Discrete Scale Invariance and log-periodic oscillations}
Let us consider a model in which the first node has no outgoing edge, while the others
have exactly one, as it happens in KR or BA models. Then, a graph with $d$ nodes has
$d-1$ edges, and for each edge $x\sim y$, $|d(x)-d(y)|\le d-2$. Thus,
$E_2(G)\le E_2^{max} := (d-1)(d-2)^2$.  This level of energy is reached with
probability $1$ in KR when $r=0$.

\begin{figure}[ht]
    \begin{center}
        \includegraphics[scale=0.5]{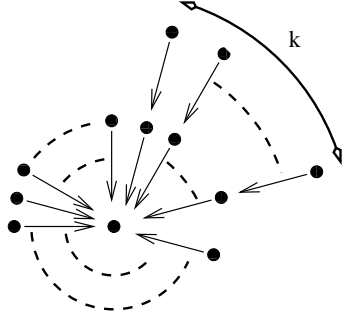}
    \end{center}
    \caption{Graph with $k$ edges in a second row}
    \label{fig:hek}
\end{figure}

Now, consider the graph $G_k$ depicted in figure \ref{fig:hek}. It has $1$ node at the
center, $k$ nodes with degree $2$, $k$ nodes ``in a second row'', attached to the
previous $k$ nodes, and $d-2k-1$ nodes with degree $1$ and at distance $1$ from the
origin.

\begin{figure}[ht]
    \begin{center}
        \includegraphics[scale=0.9]{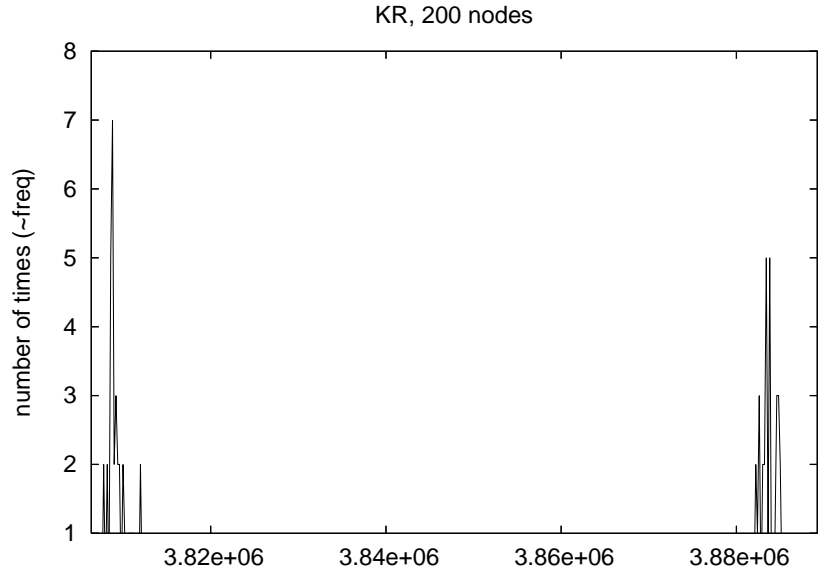}
    \end{center}
    \caption{Sub-peaks at high energies. KR, $200$ nodes, $r=0.25$, $E_2$.}
    \label{fig:sbpks2}
\end{figure}

\begin{figure}[ht]
    \begin{center}
        \includegraphics[scale=0.9]{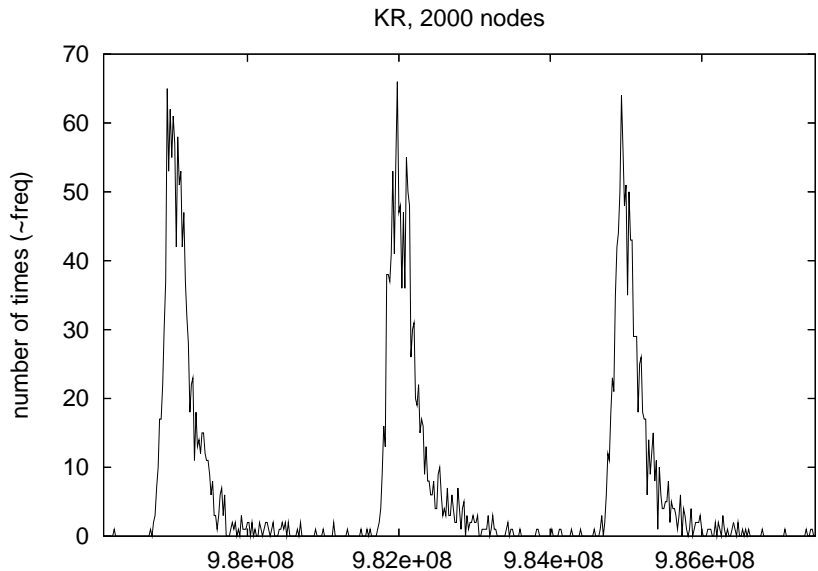}
    \end{center}
    \caption{Sub-peaks become close. KR, $2000$ nodes, $r=0.20$, $E_2$.}
    \label{fig:sbpks-cls}
\end{figure}

The energy of $G_k$ is given by
$$e_k := E_2(G_k) = (d-2k-1)(d-k-2)^2+k(d-k-3)^2+k.$$
If $k\ll d$, $$e_k\sim (d-k-1)^3.$$  Within the KR model, if $r$ is close to $0$, this
sort of graphs have non-negligible probability, and they explain the peaks at energies
close to $E_2^{max}$, which can be seen in Figure \ref{fig:sbpks2}. If one pays a
closer attention to one of these peaks, one will see that each of them has sub-peaks
with smaller probability (see Figure \ref{fig:sbpks-cls}).

These sub-peaks are due to graphs similar to $G_k$, but for which the $k$ nodes at
second row do not attach to those in first row so regularly.  The number of energy
levels around the peak associated to $G_k$ is related to the number of graphs on $k$
nodes (more precisely, to the number of \emph{forests}, i.e. disjoint union of trees,
on $k$ leaves). This means that when $k$ grows, these peaks get wider, and eventually
they become close enough to seem oscillations. Such a phenomenon can be seen in Figure
\ref{fig:sbpks-cls}.

Since, for $k \ll d$,
$$\ln(e_k) - \ln(e_{k+1}) \sim \frac{3}{d-k-2} \sim \frac{3}{d-2}, $$
the peaks can be considered a log periodic structure. The log periodicity of
oscillations was related to the existence of a {\it discrete scale invariance}
(as opposed to a continuous one), which is due to the existence of preferred scaling
ratios. For self similar fractals like the triadic Cantor set or the Sierpinski
gasket, the existence of {\it complex} fractal dimensions were proposed, since the
imaginary part leads to log periodic corrections of the scaling.

In our setting, the existence of preferred scaling ratios correspond to the existence
of different hierarchies in the distribution of node degrees in a given graph. That
is, the degrees are not evenly distributed, but lacunary.

We wish to stress that this fact was obtained independently for the degree
distribution of hierarchical scale free graphs in \cite{SH}, and some models of graph
generation were proposed. Here, the energy shows that those kind of networks are
obtained also from classical models of graphs generation like the BA, KR, or DMS.

\bigskip

\section{Conclusions}

In this work we introduced the linking energy $E_2$, and computed it for several
graphs from 50 to 100,000 nodes. We generated the graphs with preferential attachment,
edge redirection and attachment to edges models. These models are known as responsible
of power law distribution of the nodes degree, and the graphs generated are considered
scale free. We also consider random (Erd\H os-Renyi) and regular lattice graphs.

The frequency distribution of energies shows log periodic oscillations in the tail for
the graphs generated with scale free models, which are associated to complex fractal
dimensions and discrete scale invariance. They are not present for random or regular
graphs.

There are two interpretations of the terms \emph{scale free}. In the first one, the
range of the nodes degree distribution crosses several scales, instead of being
concentrated around the mean number of connections. The second one suggest that the
nodes degree distribution presents details at every interval of the range following a
power law.

Our results show the presence of discrete scale invariance in high energies graphs:
they are scale free in the first sense, although the nodes degree are not evenly
distributed between the minimum and maximum values, but clustered in hierarchies. This
is a fact observed in small networks gathered from real data. Also, for small
energies, the distribution does not show the oscillations and the graphs seems to be
scale free in the second sense. Hence, the models considered are able to generate both
kind of graphs, and the energy could be used to discriminate between them.

\end{document}